\documentclass[10pt,journal,compsoc]{IEEEtran}
\usepackage{etoolbox}
\makeatletter
\patchcmd{\@makecaption}
  {\scshape}
  {}
  {}
  {}
\makeatother
\usepackage[table]{xcolor}
\usepackage[T1]{fontenc}
\usepackage{cite}
\usepackage{amsmath,amssymb,amsfonts}
\usepackage{multirow}
\usepackage{graphicx}
\usepackage{textcomp}
\usepackage{xcolor}
\usepackage{csquotes}
\usepackage{hyperref}
\usepackage{mathtools}
\usepackage[ruled, lined, linesnumbered, commentsnumbered, longend]{algorithm2e}

\def\BibTeX{{\rm B\kern-.05em{\sc i\kern-.025em b}\kern-.08em
    T\kern-.1667em\lower.7ex\hbox{E}\kern-.125emX}}

\ifCLASSINFOpdf
\else
\fi
\usepackage{amsmath}
\usepackage[cmintegrals]{newtxmath}
\begin{document}
%
\title{A novel non-linear transformation based multi user identification algorithm for fixed text keystroke behavioral dynamics }
%
%
%

\author{Chinmay~Sahu,
        Mahesh~Banavar,~\IEEEmembership{Senior Member,~IEEE,}
        and~Stephanie~Schuckers,~\IEEEmembership{Senior Member,~IEEE}
\thanks{The authors are with the Department of Electrical and Computer Engineering, Clarkson University, Potsdam, NY, 13676 USA. This work is supported in part by the NSF CPS award 1646542.
}
}

%
%


\markboth{} 
{Sahu, \MakeLowercase{\textit{et al.}}: A novel non-linear transformation based multi-user identification algorithm for fixed text keystroke behavioral dynamics}

%



\IEEEtitleabstractindextext{\begin{abstract}
In this paper, we propose a new technique to uniquely classify and identify multiple users accessing a single application using keystroke dynamics. This problem is usually encountered when multiple users have legitimate access to shared computers and accounts, where, at times, one user can inadvertently be logged in on another user's account. Since the login processes are usually bypassed at this stage, we rely on keystroke dynamics in order to tell users apart. Our algorithm uses the quantile transform and techniques from localization to classify and identify users. 
Specifically, we use an algorithm known as ordinal Unfolding based Localization (UNLOC), which uses only ordinal data obtained from comparing distance proxies, by ``locating'' users in a reduced PCA/Kernel-PCA/t-SNE space based on their typing patterns. 
Our results are validated with the help of benchmark keystroke datasets and show that our algorithm outperforms other methods.  

\end{abstract}

}

\maketitle

\IEEEdisplaynontitleabstractindextext
%
\IEEEpeerreviewmaketitle

\section{Introduction}
\label{sec:intro}

With increasing digital presence, securing sensitive and personal data in online applications is of prime importance for maintaining privacy and security \cite{li2011relationship,wang2016targeted,ur2016users,ayotte2019fast,amrollahi2020survey,kim2020freely}. 
 In general, systems or web applications utilize one-time authentication using single sign-on for providing security.  Banking and financial institutions generally use a knowledge-based mechanism to complement the existing authentication system. However, research shows that one-time knowledge-based systems are vulnerable to attack \cite{chiasson2011persuasive}. Also, in some situations, such as where multiple users share a computer or accounts (for example, a household with a single shared computer) there is a need to not only continuously secure the system from outside attack, but also determine if wrong accounts are being accessed inadvertently.  

One solution is to authenticate users based on their behavioral biometrics  \cite{yampolskiy2008behavioural}. Behavioral biometrics involves recognizing users based on how they interact with their devices. One example is keystroke dynamics \cite{bergadano2002user,gunetti2005keystroke, hu2008k, killourhy2009comparing, messerman2011continuous, zhong2012keystroke,bours2015continuous, antal2016mobikey, sun2016shared,  murphy2017shared, dhakal2018observations, ayotte2019fast, ayotte2019Biosig}, 
where user profiles are modeled based on their typing patterns. Unlike other biometric modalities, it is easier to collect keystroke data in the background without impacting the users typing experience. Hence, keystroke-based security systems are a user-friendly and effective method of continuous user authentication.

Keystroke dynamics can be categorized as fixed-text and free-text. In fixed-text \cite{de1997enhanced,meszaros2007strengthening,killourhy2009comparing,zhong2012keystroke,antal2016mobikey,sun2016shared},   users type a specific passphrase such as a user id and a corresponding password in a login form, while free-text \cite{gunetti2005keystroke,sim2007digraphs,messerman2011continuous,dhakal2018observations,ayotte2019Biosig,ayotte2019fast,kim2020freely}  
is where users type in a free uncontrolled environment. Users can type, and therefore, create keystrokes, from either physical keyboards \cite{killourhy2009comparing} or from touch screen keyboards \cite{antal2016mobikey}. In this paper, we consider both sources of keystrokes. 

Earlier studies have mostly focused only on single-user authentication \cite{zhong2012keystroke, killourhy2009comparing, bergadano2002user}, where a profile is built for only one user. The algorithms used in single-user authentication determine whether the user at the keyboard is the user in the model. 
Extending this to multi-user identification requires  progressive testing of each user's template with the rest of the users, where prior class label information is required and implementation takes longer as the number of classes increases. 

In this paper, the goal of our work is to determine which one user among many authorized users, is at the keyboard. One example use-case is with this approach would make sense in  OTT/streaming platforms where multiple authorized users are authorized to use a single account, and have their own individual profiles within the same account. It is important to note here that since all the users are logging into the same account, they all use the same password. Currently, users have to log in first, and then select their profile to start streaming. With our proposed approach, the profile can automatically be selected based on the typing pattern of the user, which is seen as the password is entered. are present and our proposed algorithm can be added as an extra security layer to identify users and secure the platform based on users password typing pattern. 
We have demonstrated efficacy with up to 5 users demonstrating with more than 92.51\% identification accuracy. In the problem we consider here, we assume no prior knowledge of the number of users or their keystroke features. Therefore, we propose a multi-user identification system utilizing keystroke dynamics. We note here that there is very limited work in the literature on the design of multi-user classifiers based on keystroke dynamics \cite{hwang2009account}.

Another way for user authentication based on keystroke dynamics is with deep learning. In \cite{giot2019siamese}, the authors trained a Siamese network for keystroke authentication using 200 samples per user from the CMU keystroke dataset \cite{killourhy2009comparing}. In \cite{zhao2019keystroke}, a multilayer deep belief network is proposed to authenticate users from 104 typing samples.  Similarly, researchers have explored fast neural network architecture \cite{maheshwary2017deep} and convolutional neural networks \cite{cceker2017sensitivity} to authenticate users. However, all these deep learning-based models require at least 200 keystroke samples per user to train. In contrast, we use far fewer keystrokes to achieve our goal of identification based on keystrokes from a password or a short phrase. Our algorithm consists of several steps including keystroke data preprocessing using quantile transformation followed by dimensionality reduction, data clustering, and localization to identify authorized users.  The steps in our algorithm can be summarized as follows: 
\begin{itemize}
    \item We transform the statistics of the keystroke features into a uniform distribution using the quantile transform.
    \item We use dimensionality reduction techniques to extract features and project them to a reduced feature space. We compare different methods such as PCA, Kernel-PCA, and t-SNE.
    \item Unsupervised data clustering algorithms are used on the extracted features to identify the number of user clusters present in the reduced feature space. Different clustering algorithms, including DBSCAN, GMM, and X-Means, are evaluated for their effectiveness.  
    \item Finally, we adapt and use a specific indoor target localization algorithm known as ordinal unfolding-based localization (ordinal UNLOC \cite{OrdinalUNLOC}) to place the test sample in the reduced feature space, and then classify/identify the test sample.
\end{itemize}

As part of our work, we also evaluated multiple dimensionality reduction techniques and multiple unsupervised clustering techniques to identify the combination that is most suited to this application.

The rest of the paper is organized as follows.  In Section \ref{sec:related}, a brief review of the literature in keystroke dynamics is presented. 
Our algorithm method is introduced in Section \ref{sec:new_algo}. 
Results are presented in Section \ref{sec:results}. A short summary, future work, and concluding remarks are presented in Section \ref{sec:conclusions}.

\section{Related work}
\label{sec:related}

Keystroke dynamics is an active research domain. Early studies focused on authenticating users based on fixed-text such as passwords during the login process \cite{umphress1985identity}. More recently, there has been a rapid increase in interest to continuously authenticate users while they are typing \cite{messerman2011continuous}. 

The keystroke features are primarily the timing information of the key down/hold/up events. The dwell time or press time of individual keys and the flight time between two keys are typically used. Diagraphs, which quantify the latencies between two successive keystrokes for specific letter-pairs are regularly exploited in studies \cite{ayotte2019Biosig}. Similarly, $n$-graphs, which are time latencies between $n$ consecutive keys, have also been investigated. It has been shown that digraphs and $n$-graphs can be used to authenticate users with high accuracy \cite{gaines1980authentication, sim2007digraphs, messerman2011continuous}.  

The relative flight times for different $n$-graphs were used to formulate features and the results were found to be robust to intra-class variations. In \cite{gunetti2005keystroke}, the authors used trigraphs and $n$-graphs to design a relative, absolute distance based score among users to improve authentication performance in free-text. Peacock, \textit{et al.} \cite{peacock2004typing},  surveyed recent developments in keystroke authentication, compared results of several metrics, and analyzed the potential roadblocks to widespread implementation of keystroke biometrics. In \cite{araujo2005user}, the authors use key down and key up times, key press duration, and the key ASCII codes as features. Using the four features, authors achieved a false rejection rate of 1.45\% and a false acceptance rate of 1.89\% for user authentication. Meszaros, \textit{et al.} \cite{meszaros2007strengthening}, designed an adaptive distance-based threshold on features to authenticate users. Harun, \textit{et al.} \cite{harun2010performance}, proposed a multilayer perceptron neural network with a backpropagation learning algorithm as a classifier.   Similarly, many machine learning and identification techniques based on methods such as  distance \cite{killourhy2009comparing,ayotte2019Biosig}, Bayesian classifiers  \cite{messerman2011continuous}, SVM \cite{abramson2013user}, fuzzy logic \cite{de1997enhanced}, neural networks  \cite{li2001remote}, K-nearest neighbor \cite{hu2008k}, and K-means  \cite{kang2007continual} have been exploited in keystroke biometric research \cite{killourhy2009comparing,li2011study}.

In recent literature \cite{hitaj2019passgan,kang2015keystroke,ali2017keystroke,roth2014continuous,ayotte2020fast,krishnamoorthy2018identification}, Keystroke research focused on fixed text \cite{hitaj2019passgan,ali2017keystroke,krishnamoorthy2018identification} and free text \cite{kang2015keystroke,roth2014continuous,ayotte2020fast}. Earlier studies in literature focused only on single-user authentication or free text-based authentication. In single-user authentication, the model is built for one user only. The algorithms are used to determine whether the user at the keyboard is the user in the model or not. However, in this paper, we focus on multi-user identification. We would also like to point out that this type of work is relatively unaddressed in literature, and we were only able to find one paper addressing this very important issue \cite{hwang2009account}. We would also like to note that relatively less work has been done with Clarkson Account Recovery Dataset \cite{wahab2022securing}, and therefore, we will not discuss it here, but leave it for future exploration.   

All the algorithms except Hwang, \textit{et al.} \cite{hwang2009account} are for single user authentication.
Here, the authors used a variational-based Gaussian mixture model (VB-GMM) on raw keystroke features to determine whether the account is shared or not. They used a dataset consisting of 25 passwords from 16 users to test their model.  However, profiling multiple users based on their keystrokes in a single system or application is still an open problem.  Though  Hwang, \textit{et al.} \cite{hwang2009account} presented an algorithm to identify multiple users, the algorithm is not tested on available benchmark datasets. 
  
To solve the multi-user classification problem, in this paper, we introduce a new algorithm that leverages concepts from data preprocessing, dimensionality reduction, data clustering, data embedding, and data localization, and can be directly applied to the raw keystroke data. 
To our knowledge, this is the first time such an approach has been used in keystroke dynamics.
We discuss the algorithm in detail in Section \ref{sec:new_algo}, and demonstrate its effectiveness using two publicly available datasets in Section \ref{sec:results}. 

\section{Proposed Algorithm}
\label{sec:new_algo}

In what follows, we present an algorithm to identify multiple users accessing a single system or application based on their typing patterns.  The proposed multi-user identification algorithm can be split into  training and test phases. These are described below. 

\subsection{Training}
\label{ssec:training}
The keystroke data is split into training and testing sets. The set of keystroke features, $\mathbf{F}_{\rm train}$, are inputs to the training algorithm. A quantile transformation (Q-transform) is applied on the training set to transform the samples to a uniform distribution. Based on the original data distribution, the quantile transform will vary. Therefore, the specific quantile transform derived here, $Q_{\rm train}(\cdot)$, will be stored and used in the testing phase. The transformed training set, $\tilde{\mathbf{F}}_{\rm train}$, is then projected to a reduced feature space by using dimensionality reduction techniques to give a new set of features, $\tilde{\mathbf{f}}_{\rm train}$. Then, the number of user clusters, $N$, is estimated using unsupervised clustering algorithms such as DBSCAN, GMM, and X-means. Then the cluster centroid of each user feature cloud, $C_{l}, l = 1, 2, \dots, N$ is estimated and is designated as the representative point for that user. In localization-terms, this point is called the anchor and will be used in the testing stage as described in Section \ref{ssec:testing}. 

\begin{algorithm}

    \KwIn{A set of incoming keystroke features: $\mathbf{F}_{\rm train}$ }
    \KwOut{ Quantile transformed function: $Q_{\rm train}(\cdot)$, \\ \hspace{37pt} Quantile transformed data: $\tilde{\mathbf{F}}_{\rm train}$, \\ \hspace{37pt} Features in reduced feature space: $\tilde{\mathbf{f}}_{\rm train}$,\\\hspace{37pt} Number of cluster centroids: $N$,
    \\\hspace{37pt} Cluster centroids: $C_{l}$ where $l = 1, 2, \dots, N$.}
Apply Q-transform on $\mathbf{F}_{\rm train}$ to generate $Q_{\rm train}(\cdot)$, $\tilde{\mathbf{F}}_{\rm train}$;

Apply PCA/ K-PCA/ t-SNE on  $\tilde{\mathbf{F}}_{\rm train}$ to generate $\tilde{\mathbf{f}}_{\rm train}$; 

Apply DBSCAN/ GMM/ X-means on $\tilde{\mathbf{f}}_{\rm train}$ to obtain user clusters;

Estimate cluster centroid $C_{l}$;

Return $Q_{\rm train}(\cdot)$, $\tilde{\mathbf{F}}_{\rm train}$, $\tilde{\mathbf{f}}_{\rm train}$, $N$, $C_{l}$.
    \caption{Training}
    \label{algo:training}
\end{algorithm}

\vspace*{-.5cm}

\subsection{Testing}
\label{ssec:testing}
\begin{figure}
\centering
\includegraphics[width=0.4\textwidth]{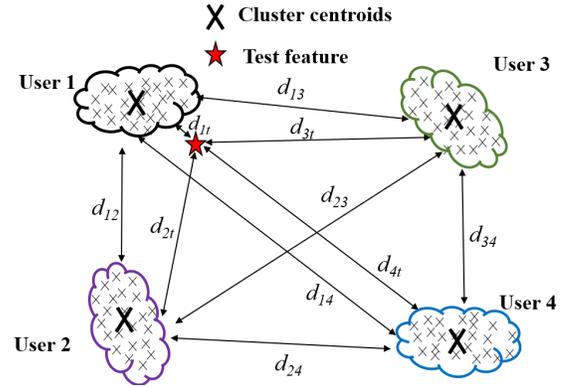}
 \caption{A visual example of the proposed localization. Reduced feature space that contains user feature clouds (users at known locations), anchors (cluster centroid of the features), and unidentified user test feature  (users at unknown locations, marked as the test feature). Only comparative pairwise distances are reliable, giving rise to ordinal data.}
 \label{fig:locafigure}
\end{figure}

Keystroke data samples from the testing set are used to identify which one of the many trained users they belongs to. Consider the keystroke sample from a test user (TU), $F_{\rm test}$ from the set of training data, $\mathbf{F}_{\rm test}$. This sample is transformed by the same quantile transformation function, $Q_{\rm train}(\cdot)$, generated during the training process to give us $\tilde{F}_{\rm}$. 
Using all keystroke data samples $\tilde{\mathbf{F}}_{\rm train}$  from each of the known users (KU), two types of cross-distances are computed using the scaled Manhattan distance. These are: known user to known user distance, $d_{lm}$, where $l$ and $m$ are known user samples, and known user to test user cross-distance, $d_{mt}$, where $m$ is a known user and $t$ is the test user (see Figure \ref{fig:locafigure}). These distances are then compiled into a cross distance matrix, $\mathbf{d}$.  

\begin{algorithm}

    \KwIn{A keystroke sample: $F_{\rm test}$,\\ 
    \hspace{30pt}Quantile transformed function : $Q_{\rm train}(\cdot)$, \\
    \hspace{30pt}Quantile transformed data: $\tilde{\mathbf{F}}_{\rm train}$, \\
    \hspace{30pt}Reduced feature space: $\tilde{\mathbf{f}}_{\rm train}$,\\
    \hspace{30pt}Number of cluster centroids: $N$,\\
    \hspace{30pt}Cluster centroids: $C_{l}$
    }
    \KwOut{A predicted or assigned user cluster label: $\hat{C}_l$}
Fit $F_{\rm test}$ using $Q_{\rm train}(\cdot)$ to obtain $\tilde{F}_{\rm}$ ;

Estimate cross-distances $d_{lm}$ and $d_{mt}$;

Estimate cross distance matrix $\mathbf{d}$;

Apply Ordinal UNLOC using $\mathbf{d}$ and $\tilde{\mathbf{f}}_{\rm train}$  to estimate test feature location:  $\tilde{f}_{\rm test}$;

Apply knn using $\tilde{f}_{\rm test}$  to assign a user label: $\hat{C}_l$.

    \caption{Testing  }
\end{algorithm}

Ordinal UNLOC \cite{banavar2021ordinal} is then used to perform two tasks: (1) the keystroke data is mapped from the keystroke space to the reduced dimension space using function mapping; and (2) the ``location'' of the unknown user in the reduced feature space is estimated. Using the nearest neighbor rule, this estimate is mapped to one of the known users, and a classification decision is made. 

We should note here that the function learning step in Ordinal UNLOC is crucial, since it maps data from keystroke space to reduced dimensions, without resorting to the computationally expensive dimensionality reduction process. Further, as will be seen from the simulations, Ordinal UNLOC, since it is designed to function with noisy and incomplete data, is ideally suited to keystroke data, and improves performance over conventional dimensionality reduction techniques. 

\begin{figure*}
\centering
\includegraphics[width=0.8\textwidth]{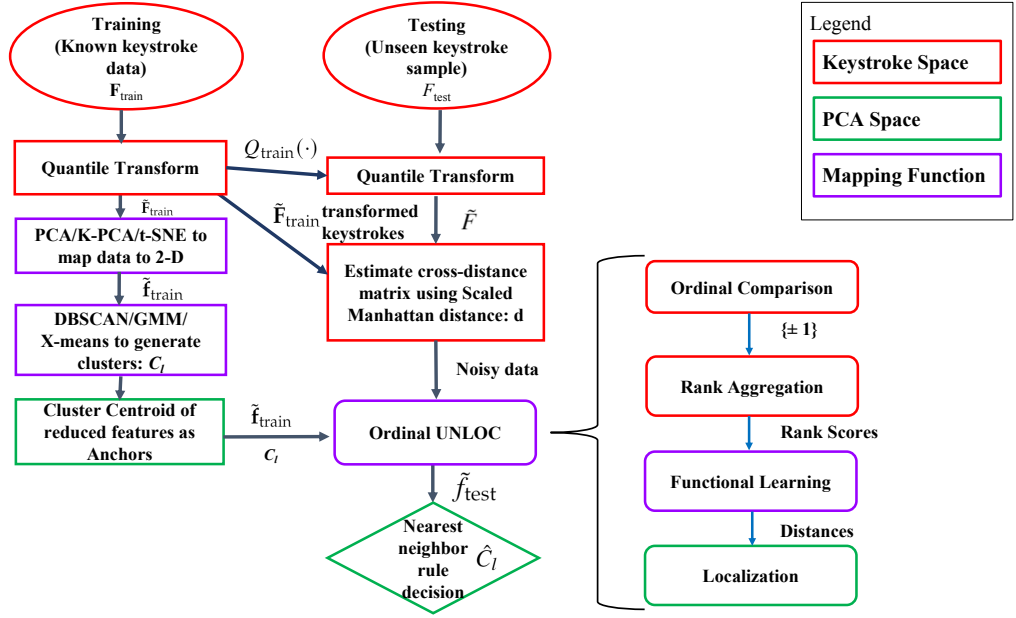}
 \caption{Flow chart of the proposed algorithm to estimate the location of users' based on ordinal data. The method uses distances estimated directly from keystroke features in reduced feature space with Ordinal UNLOC algorithm to localize test features.}
 \label{fig:flowchart}
\end{figure*} 

\section{Methods}
\label{ssec:methods}

In what follows, we describe the feature transformation method we use, as well as the scaled Manhattan distance, the dimensionality reduction techniques, and the clustering algorithms. 

\subsection{Feature Transformation}
\label{sssec:quntile tranformation}
Since the keystroke typing timestamps collected are prone to variances due to user typing behavior, there always will be some outliers present in the incoming keystroke data. Since all users are typing the same text, some samples from different users may be too close to each other, leading to misclassification. To address these problems, we implement a quantile-based transformation (Q-transform) \cite{hilger2006quantile} on the incoming features to convert the user features into a uniform distribution. Hence, the outliers get pushed closer to the corresponding user features, and features that are close to each other, are spaced out. 
 
The quantile-based transform is a non-linear transformation technique that is based on the monotonic transformation of features. It transforms all the features into the desired distribution based on a function, $F^{-1}[\phi(\cdot)]$, where $\phi(\cdot)$ is the cumulative distributive function of a feature and $F^{-1}$ is the quantile function of the desired output distribution $F$. 
We select the desired quantile function to be uniform distribution so that outliers are bought closer to the rest of the data, and closely spaced data get a little separation. 

Since the Q-transform is specific to $\phi(\cdot)$, the function derived during training is stored, and this same function is used throughout the rest of the process. 

\subsection{Scaled Manhattan Distance}
\label{sssec:manhattan}
The Scaled Manhattan distance is derived from the Manhattan distance. Consider a set of features with $N$ samples of the $k$-th user, placed in $l$-dimensional space, given by $\mathbf{f}_{i,j,k}$, $i = 1, 2, \dots l, j = 1, 2, ... N$. Let $\mathbf{m}_{i,k}$ be the mean of $i$-th feature across the $N$ samples, and $\sigma_{i,k}$ be the absolute deviation across the $N$ samples of the $k$-th user relative to which the distance score is estimated. The Manhattan distance (popularly known as the L-1 distance or the city-block distance) between the point $\mathbf{f}_{i,j,k}$ and the feature mean $\mathbf{m}_{i,k}$ is  defined as $||\mathbf{f}_{i,j,k}-\mathbf{m}_{i,k}||_1$. The Scaled Manhattan distance of the $k$-th user relative to a test feature $\mathbf{f}_{i,j,\tilde{k}}$ is estimated as
\begin{equation}
d^{{\rm (SM)}}_{k,\tilde{k}} =  \frac{1} {l} \sum_{i=1}^{l} \frac{||\mathbf{f}_{i,j,\tilde{k}}-\mathbf{m}_{i,k}||_1} {\sigma_{i,k}},
\label{eqn:scaled_manhattan}    
\end{equation}
and gives us a dissimilarity score. Note that while $\mathbf{f}$ and $\mathbf{m}$ can be in higher dimensional space with the same dimensions, the L1 distance between them is a scalar value. Since, in this paper, we consider only the scaled Manhattan distance, we drop the superscript and denote the distance measure as $d_{k,\tilde{k}}$.

\subsection{Dimensionality Reduction Techniques}
\label{sssec:dim_reduction}
To represent $l$-dimensional keystroke features in reduced 2-D feature space, dimensionality reduction techniques are applied.  PCA, Kernel-PCA, and t-SNE are explored as options for use with keystroke features in our algorithm.

\subsubsection{Principal Component Analysis (PCA)}
\label{ssssec:PCA}
Principal component analysis (PCA)  is a dimensionality reduction technique that is generally used to condense high-dimensional data to a lower dimension to simplify interpretation. The basis of PCA is to transform the original multi-dimensional dataset to a lower dimension without losing the high variations within the data \cite{abdi2010principal}. In literature, PCA is shown to be used as a preprocessing step for feature reduction, to simplify analysis, and improve the accuracy of algorithms \cite{shaker2014keystroke}. In this paper, PCA is applied to raw keystroke features and quantile transformed features to obtain reduced feature spaces as shown in Figure \ref{fig:quantile transformation} and \ref{fig:quantile_transformation_mobiKey}.

Kernel PCA is a non-linear extension of PCA, for cases where linear PCA may not capture and separate features as expected  \cite{mika1999kernel}. Kernel PCA maps the data into some high dimensional space using a non-linear function and then performs linear PCA on the mapped data. In this paper, a radial basis function is used as the kernel for projecting features, which improves the separation of user features in reduced feature space.

\subsubsection{t-Distributed Stochastic Neighbor Embedding (t-SNE)}
\label{para:tSNE}

t-Distributed Stochastic Neighbor Embedding (t-SNE) is a non-linear technique for dimensionality reduction that is well suited for the visualization of high-dimensional datasets \cite{van2008visualizing}. t-SNE converts Euclidean distances between datapoints in higher dimensions into conditional probabilities that represent similarities. The similarity of datapoints, $x_i$ and $x_j$, is the conditional probability, $p_{i|j}$, that $x_j$ would pick $x_i$ as its neighbor if neighbors were picked in proportion to their probability density under a Gaussian centered at $x_j$. Note that for a pair of ppints, $x_{i}$ and $x_{j}$, due to symmetry, $p_{i|j} = p_{j|i}$. 

The next step of t-SNE is to estimate a set of datapoints in a lower dimensional space where the datapoints follow similar distribution as in the higher dimensional space. For lower dimensional datapoints, it is possible to estimate similar conditional probabilities $q_{i|j}$ for points $y_{i}$ and $y_{j}$. The algorithm  then attempts to construct a 2-dimensional embedding that minimizes the Kullback-Leibler divergence \cite{hershey2007approximating} between the vector of similarities between pairs of points in the original higher-dimensional dataset and the similarities between pairs of points in the lower embedding.

Unlike the PCA or kernel-PCA, t-SNE does not learn a mapping function to project a test input to the existing map. One drawback of the t-SNE approach is that since there is no mapping to be learned, t-SNE has to be rerun each time there is a change in the data, however small, leading to increased complexity.   

\subsection{Clustering}
\label{subsubsec:clustering}

Once we have the data in reduced feature space, the number of clusters in the data is to be estimated. We first need to estimate the number of users from the collected data. Therefore, instead of supervised algorithms such as K-means clustering that require us to provide the number of clusters, there is a need for unsupervised clustering. We evaluate Density-Based Spatial Clustering of Applications with Noise (DBSCAN) \cite{birant2007st}, the Gaussian mixture model (GMM) \cite{bouman1997cluster}, X-Means for  clustering. 

\subsubsection{DBSCAN}
\label{para:dbscan}
DBSCAN is a clustering algorithm, which finds core samples of high density and expands clusters from them. It is suitable for data which contains clusters of similar density. In DBSCAN, a transformed feature belongs to a cluster if it is close to many features from the same cluster. DBSCAN can find arbitrary shaped clusters and clusters with outliers or noise. DBSCAN is applied to CMU and MOBIKEY dataset to identify number of clusters from incoming features.

\subsubsection{GMM}
\label{para:gmm}
Gaussian Mixture Models (GMMs) are expectation-maximization-based probabilistic models that use soft clustering for grouping points in different clusters. GMMs group data points by assuming a certain number of Gaussian distributions in the data and each of those distributions represent a cluster.  

\subsubsection{X-means}
\label{para:xmeans}
X-means clustering is a variation of k-means clustering that consolidates cluster assignments by continuously forcing subdivision and preserving the best resulting splits by optimizing criteria such as Bayesian Information Criterion (BIC) or Akaike Information Criterion (AIC). Here, we use BIC to assess the current state of clustering and recommend the optimal number of components to use in GMMs and X-means. BIC is defined as 
\begin{equation}
    \rm BIC= -2*\log(L) + \log(n)*q
\end{equation}
where $L$ is the likelihood function, $n$ is the number of data points, and $q$ is the number of parameters in the model. The algorithm  then selects the model with smallest BIC score, giving us a way to choose between different models with varying numbers of parameters. We use the log of the Bayesian Information Criterion (BIC) score to find the optimal number of clusters in the data.   

The advantage of using X-means is that it is deterministic and does not require the knowledge of the number of clusters to initialize. However, it is sensitive to outliers. Therefore, when used along with quantile transformation, which minimizes the effects of the outliers, the sensitivity of X-means to outliers is reduced. 

Once we identify the optimal number of clusters, the nearest neighbor rule is applied to the reduced feature space to identify users. Overall accuracy is used to evaluate the performance of each algorithm.

\subsection{Ordinal Unfolding based Localization}
\label{subsubsec:loca}

Ordinal Unfolding based localization (in short, Ordinal UNLOC) \cite{banavar2021ordinal} is a target localization technique where we do not have reliable distance measurements between components in the system (nodes, transmit-receive pairs, etc.). The estimation technique utilizes rank aggregation, function learning, and proximity-based unfolding optimization, and as a result, it yields accurate target localization for common transmission models with unknown parameters and noisy observations that are reminiscent of practical settings \cite{OrdinalUNLOC}. The computational estimation approach consists of three main steps. First, from the ordinal comparison data, rank aggregation is applied to obtain a set of ``dissimilarities'', which, for a given reference sensor $k$, assigns a score to each sensor that serves as proxy to its distance to $k$ according to the ordinal data and rankings. Such sets of spatial proximities (scores) are not unique since any shifting and (positive) scaling preserves the ranking of the scores. Consequently, the scores obtained in this step cannot be directly used as (approximate) anchor-to-target distances for localization. Next, the estimated distance proximities among the anchors together with the (known) anchor-to-anchor distances are used to fit functions that best transform proximities into distances, and such functions are then applied to obtain estimates of anchor-to-target distances. Finally, given the location of anchors and (estimated) anchor-to-target distances, a multidimensional unfolding optimization problem is formulated for each target, whose solution represents a (best) estimation of the location of that target. 

Ordinal UNLOC utilizes noisy and incomplete data, due to the complex mutlipath profiles in indoor environments. Using this data directly in localization algorithms can lead to misleading localization results. Rather than using this data directly, Ordinal UNLOC compares  measurements to generate ordinal data. Similarly, keystroke data is noisy since users tend have variations in their typing speeds even within short observation windows. Here, again, we compare the dissimilarities between typing samples, leading to ordinal comparisons among users, and then using the framework of Ordinal UNLOC to solve the classification problem. Therefore,  Ordinal UNLOC allows for a robust approach to solve the multi-user classification problem when dealing with inherently noisy keystroke data.

\section{Numerical Results}
\label{sec:results}

Our algorithms are evaluated on the CMU keystroke \cite{killourhy2009comparing} and MOBIKEY dataset \cite{antal2016mobikey}. The keystroke data is divided into training and testing. The quantile transformation is then applied to the training data. Effectiveness of quantile transformation and dimensional reduction techniques (PCA, Kernel-PCA, and t-SNE) is assessed. Then clustering schemes are applied to the reduced features obtained from the dimensionality reduction process. The effectiveness of clustering techniques such as DBSCAN, GMM and X-means is evaluated. The accuracy of identification is then estimated for each approach for different sample sizes and different multi user cases. 

\subsection{Keystroke Datasets}
\label{ssec:keystroke}

We use two datasets for validating the proposed algorithm: (i) CMU Keystroke benchmark dataset \cite{killourhy2009comparing}; and (ii) MOBIKEY Keystroke Dynamics Password Database \cite{antal2016mobikey}.

The CMU keystroke benchmark dataset consists of static keystroke typing times from 51 subjects, each typing a password ``.tie5Roanl''  50 times over 8 sessions \cite{killourhy2009comparing}. Various timing features such as the key press times and hold times, were extracted from the raw data. These keystroke timings are used as inputs to our algorithm.

The MOBIKEY dataset \cite{antal2016mobikey} consists of users typing the same password as in the CMU dataset, ``.tie5Roanl'',  20 times over 3 sessions each one week apart. To collect this data, users were asked to type the password using the touchscreen of a Nexus 7 tablet. The authors \cite{antal2016mobikey} used a custom application to store the time, touch, and accelerometer-related raw data for each typing sessions.  The raw data consists of the point of touch down and touch up events initiated by users.  The various timing features used by researchers (e.g., the touch key pressing up-down times and hold times) were extracted from the raw data. We used up-down keystroke timings as the input features for our research.

In all our experiments that we present below, we split the data into training and testing sets in an 80:20 ratio. 

\subsection{Evaluating the Quantile transformation}
\label{ssec:effect_quantile}

First, the effectiveness of the quantile transformation is evaluated on different sample sizes. As discussed earlier in Section \ref{sssec:quntile tranformation}, keystroke features are transformed to follow a uniform distribution. This spreads out the most frequent values of keystroke features and reduces the impact of marginal outliers.  

\begin{figure}[!ht]
\begin{minipage}[b]{0.48\linewidth}
  \centering
  \centerline{\includegraphics[width=4.0cm]{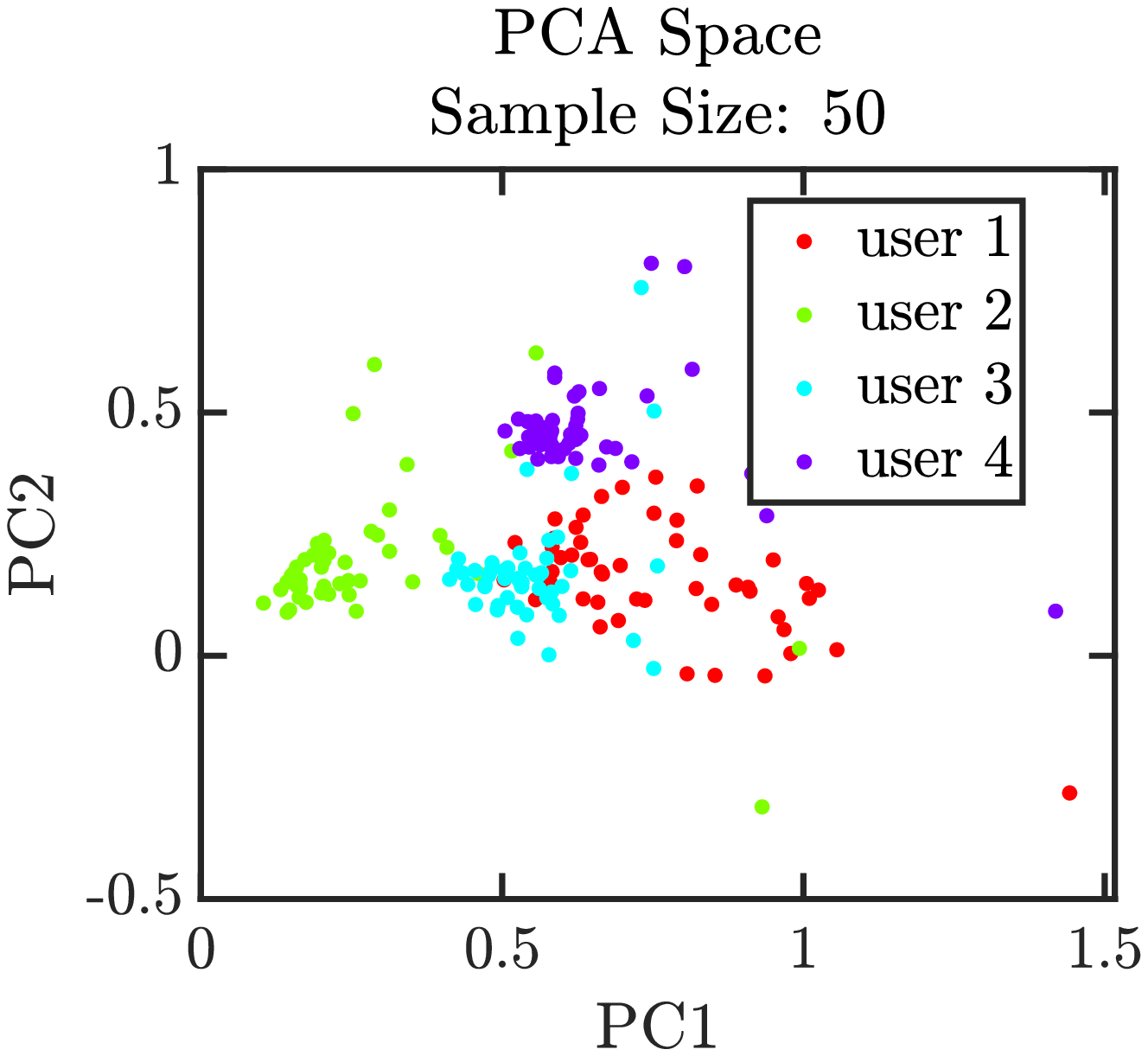}}
  \centerline{(a) PCA on raw features}\medskip
\end{minipage}
\begin{minipage}[b]{0.48\linewidth}
  \centering
  \centerline{\includegraphics[width=4.0cm]{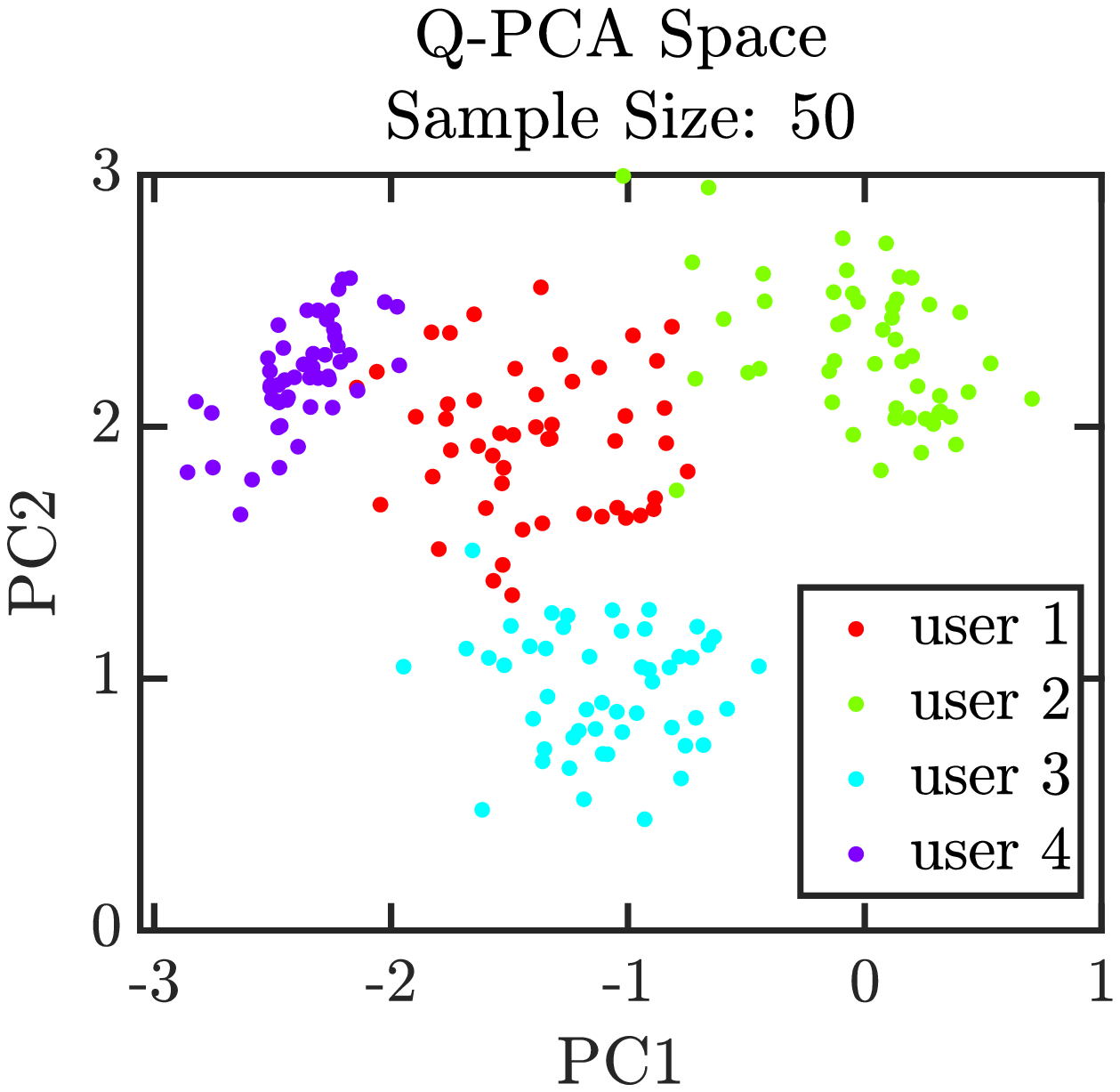}}
  \centerline{(b) PCA on transformed features}\medskip
\end{minipage}
\begin{minipage}[b]{.48\linewidth}
  \centering
  \centerline{\includegraphics[width=4.0cm]{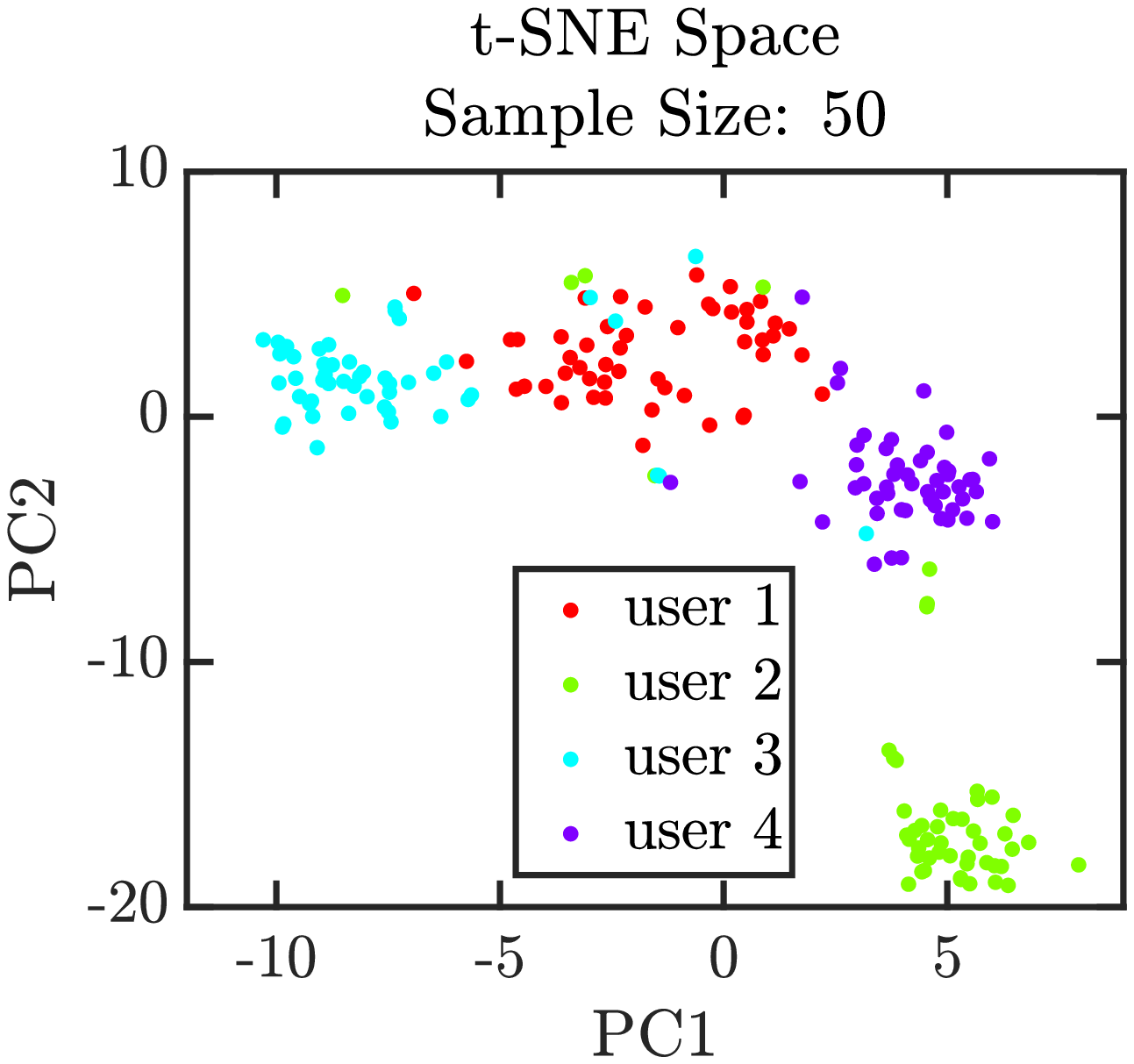}}
  \centerline{(c) t-SNE on raw features}\medskip
\end{minipage}
\hfill
\begin{minipage}[b]{0.48\linewidth}
  \centering
  \centerline{\includegraphics[width=4.0cm]{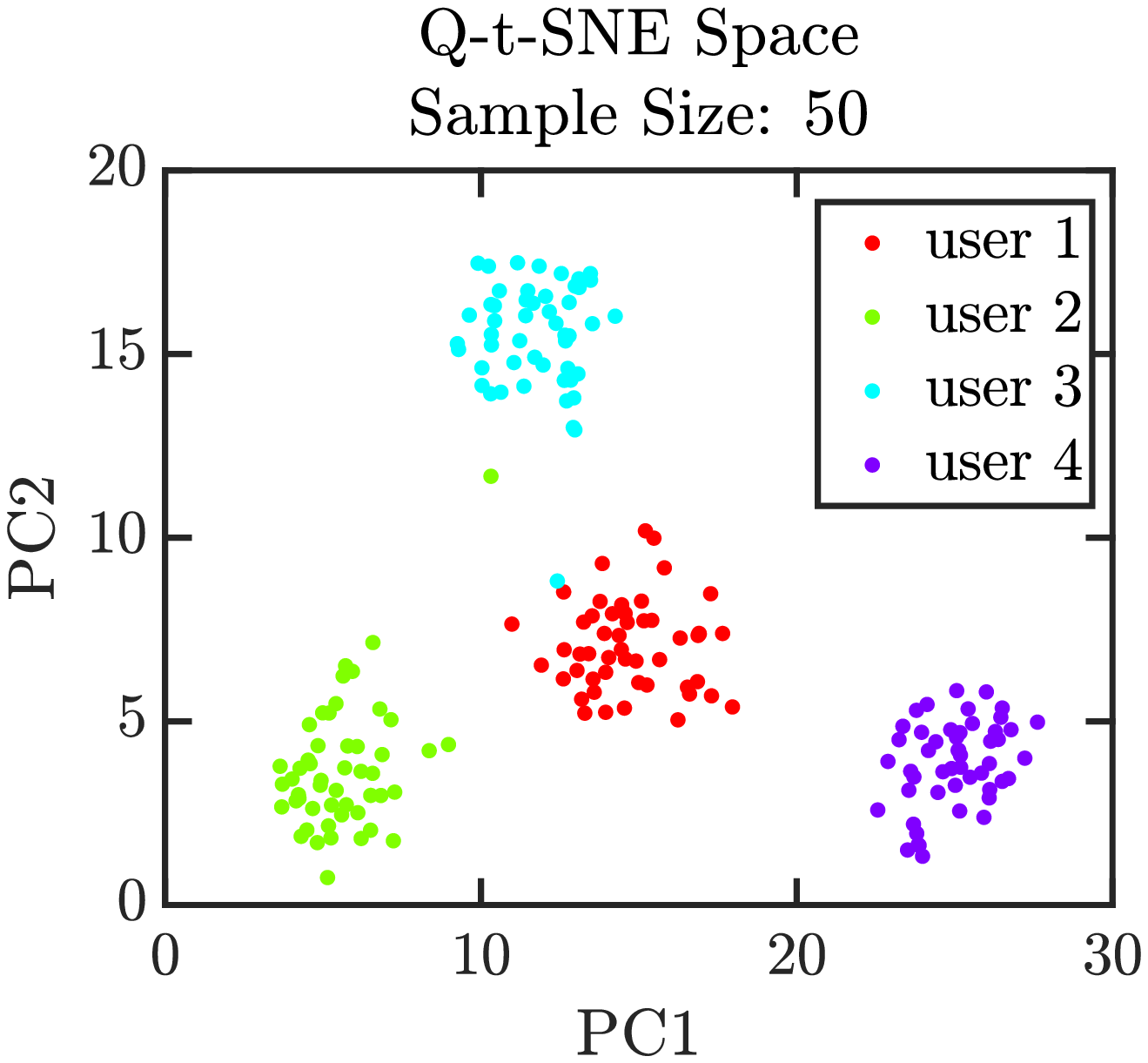}}
  \centerline{(d) t-SNE on transformed features}\medskip
\end{minipage}
\caption{PCA and t-SNE are applied to the features of CMU dataset. For both PCA and t-SNE, applying the quantile transformation visibly improves clustering.}
\label{fig:quantile transformation}
\end{figure}

\begin{figure}[!ht]
\begin{minipage}[b]{0.48\linewidth}
  \centering
  \centerline{\includegraphics[width=4.0cm]{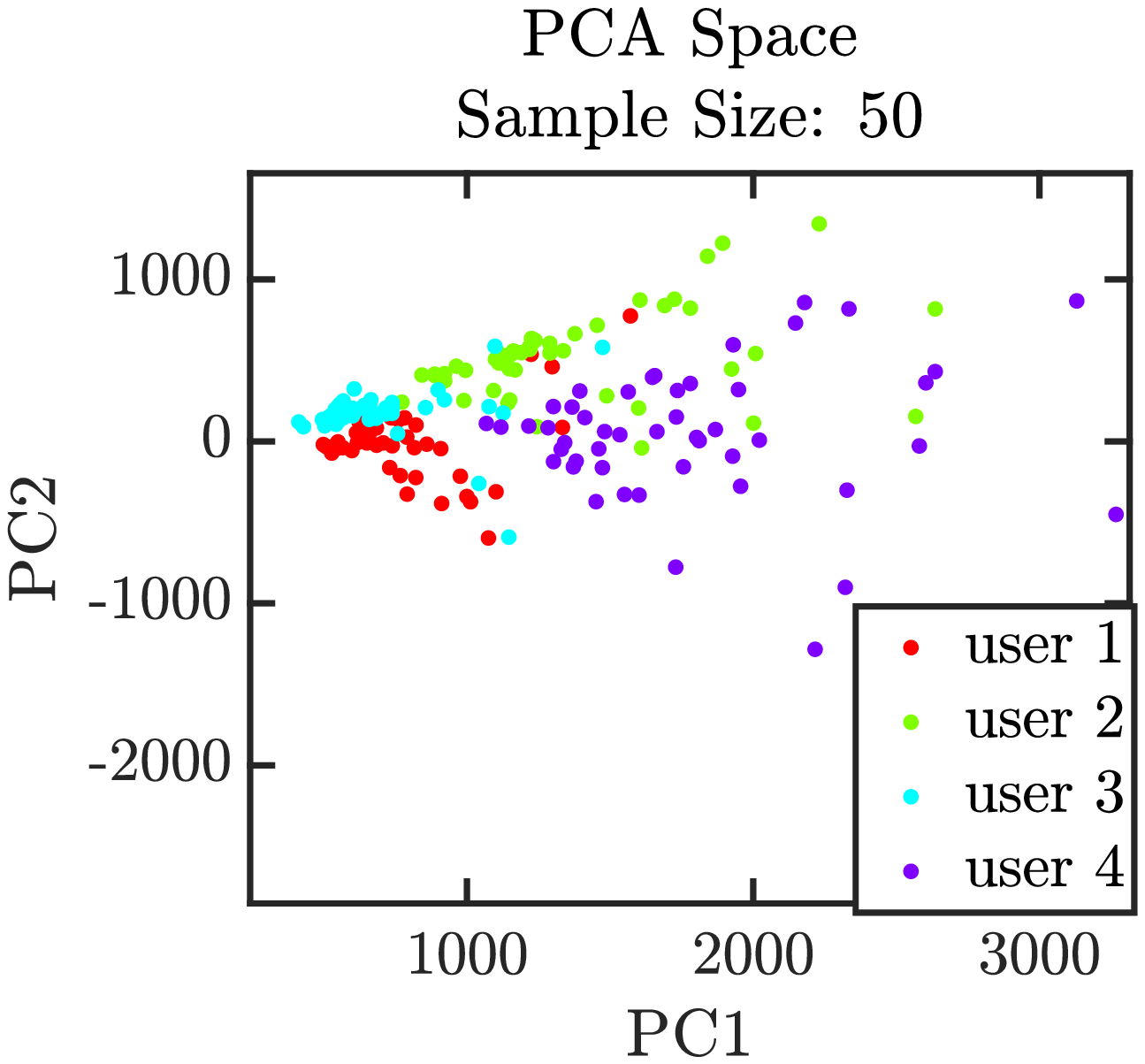}}
  \centerline{(a) PCA on digraphs}\medskip
\end{minipage}
\begin{minipage}[b]{0.48\linewidth}
  \centering
  \centerline{\includegraphics[width=4.0cm]{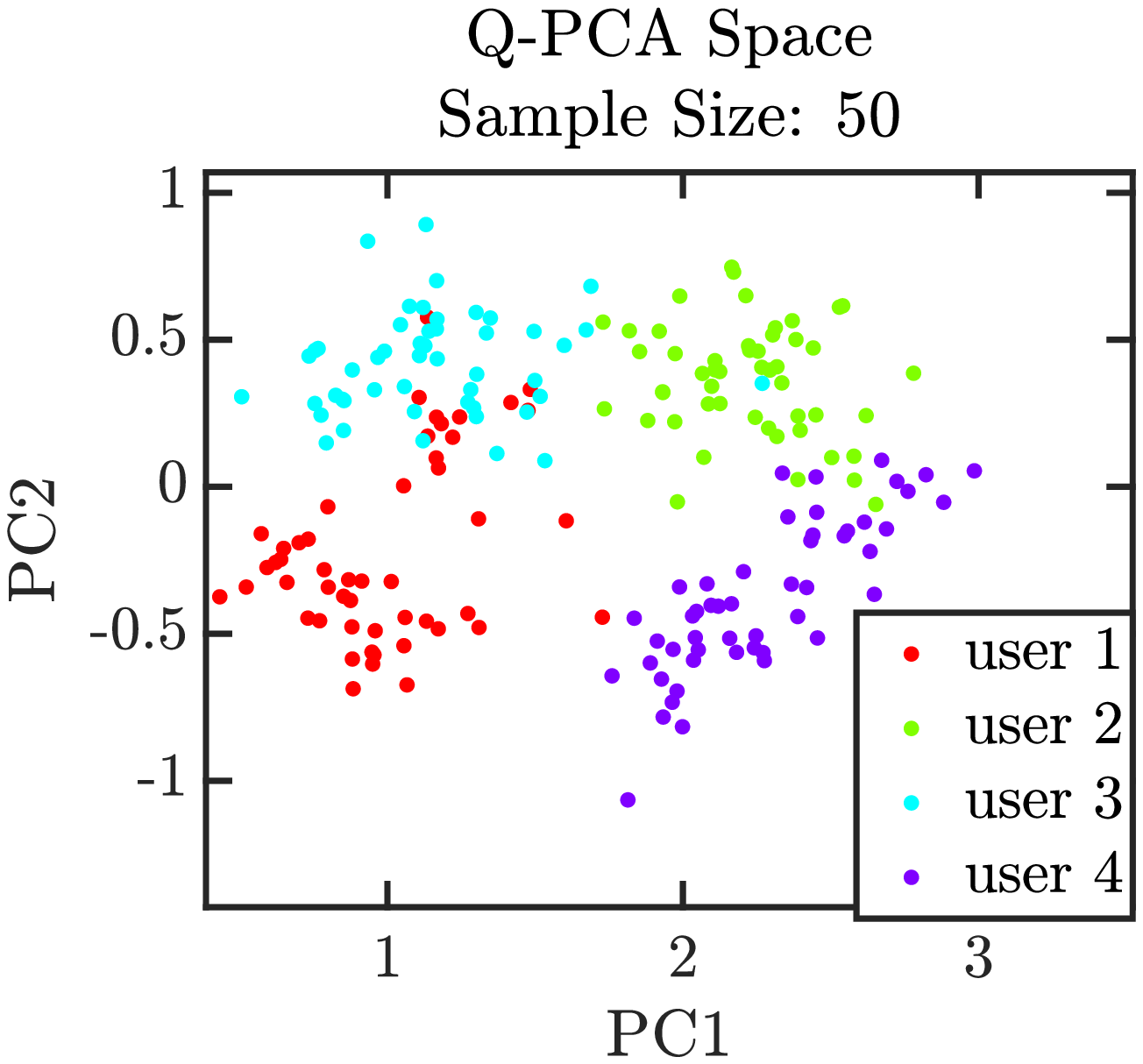}}
  \centerline{(b) PCA on transformed digraphs}\medskip
\end{minipage}
\begin{minipage}[b]{.48\linewidth}
  \centering
  \centerline{\includegraphics[width=4.0cm]{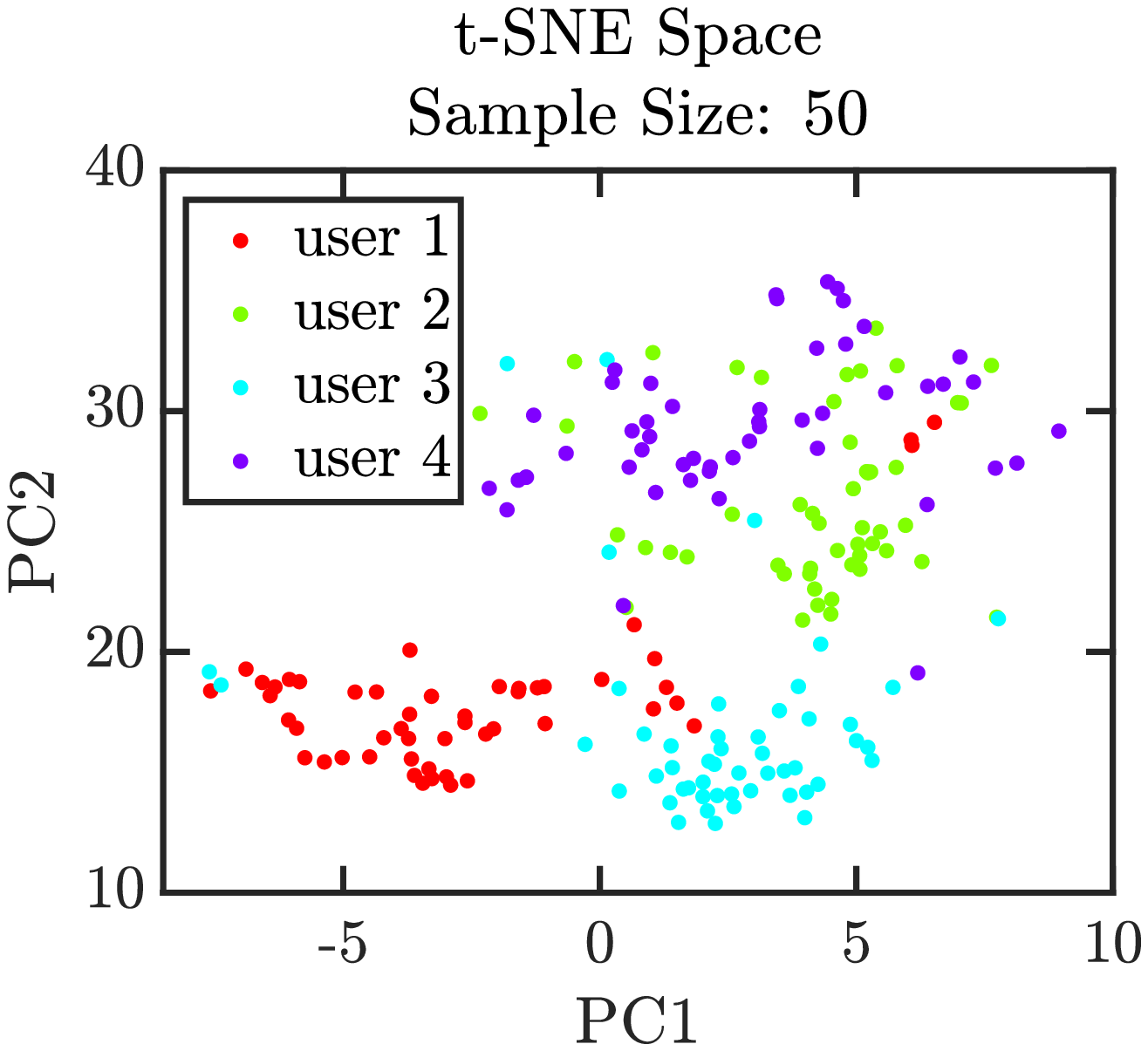}}
  \centerline{(c) t-SNE on digraphs}\medskip
\end{minipage}
\hfill
\begin{minipage}[b]{0.48\linewidth}
  \centering
  \centerline{\includegraphics[width=4.0cm]{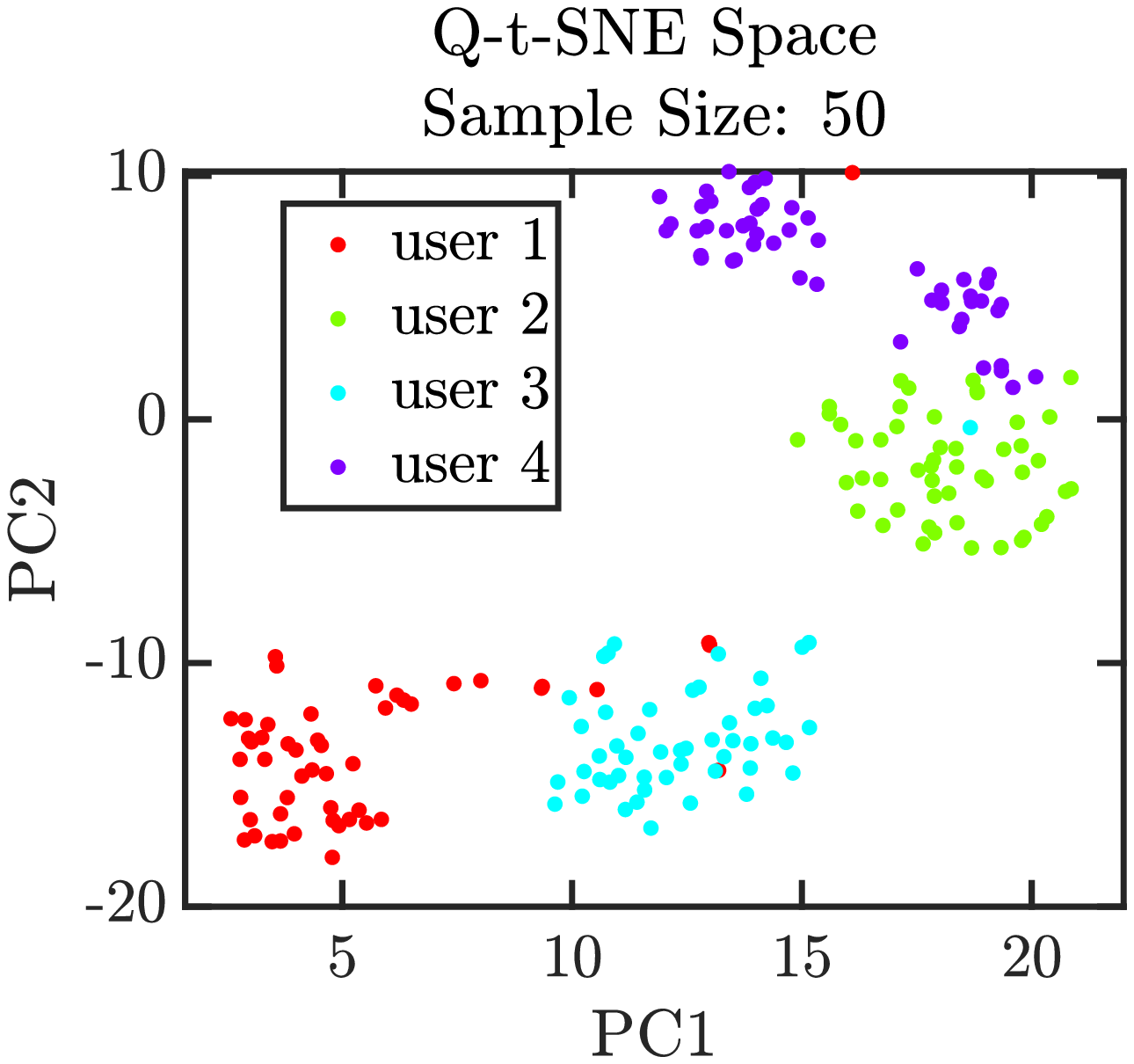}}
  \centerline{(d) t-SNE on transformed digraphs}\medskip
\end{minipage}
\caption{PCA and t-SNE are applied to the features of MOBIKEY password dataset. For both PCA and t-SNE, applying the quantile transformation visibly improves clustering.}
\label{fig:quantile_transformation_mobiKey}
\end{figure}
The quantile transformation is applied to the training feature set and the transformer function that maps the features to a uniform distribution, $Q_{\rm train}(\cdot)$, is learned. This transformer function is then used to transform test features from the same user. Then, PCA, kernel-PCA, and t-SNE are applied to both the raw and the transformed training features, resulting in projecting the features into a reduced feature space. The effects of quantile transformation on both datasets are shown in Figure \ref{fig:quantile transformation} and Figure \ref{fig:quantile_transformation_mobiKey}, where for a sample size of 50, PCA and t-SNE are applied to both the raw features and the quantile transformed features. When PCA and t-SNE are applied to raw features, we do not observe a good separation of data. However, there is a visible improvement in the user feature separation on the quantile transformed data (see Figure \ref{fig:quantile transformation} and Figure \ref{fig:quantile_transformation_mobiKey}).

\subsection{Comparison of clustering schemes}
\label{ssec:compare_clustering}

It is essential to identify the number of distinct users present in the reduced feature space. Hence, after applying dimensionality reduction techniques (PCA, Kernel-PCA, and t-SNE), unsupervised clustering algorithms such as DBSCAN, GMM, and X-means are used to identify the number of user clusters present in the keystroke data. These clustering algorithms are evaluated quantitatively based on the adjusted rand index (ARI) given  by
\begin{equation}
    \text{ARI} = \frac{\text{RI} - \text{ERI}}{\max(\text{RI}) -  \text{ERI}}, 
    \label{eqn:ARI}
\end{equation}
where RI is the rand index and ERI is the expected rand index. RI estimates a similarity measure among two clusters by analyzing all pairs of samples, enumerating the numbers of pairs, both within the same cluster, or both in different clusters (a total of four combinations), and calculating the ratio of number of pairs in the same clusters to the total number of pairs. The RI is calculated for each trial, and the max(RI) is the largest RI from this set, and the ERI is the average of the all the RI values obtained.  An ARI value of zero indicates random labeling and poor clustering. ARI values close to 1 indicate efficient clustering. 

DBSCAN is applied to the reduced feature space obtained from PCA, Kernel PCA, and t-SNE.  When using DBSCAN, care must be taken to select the correct values of two parameters: (i) the maximum distance between two samples for one to be considered as in the neighborhood of the other; and (ii) the number of samples, or total weight, in a neighborhood for a point to be considered as part of a cluster or a core point. These parameters, appear in \texttt{scikitlearn} \cite{tran2013revised} as \texttt{eps} and \texttt{min\_samples}.  

\begin{table*}[!htb]
\caption{ DBSCAN, GMM, X-Means are used for clustering (CMU). K stands for an identified number of clusters. ARI stands for the adjusted rand index. X-means predicts the number of clusters successfully in quantile transformed space. Larger ARI values in the table below show that X-means outperforms both DBSCAN and GMM. Similar results were seen for the MOBIKEY dataset.}
\label{table:DBSCAN_GMM_Xmeans}
\centering
\begin{tabular}{|c|c|cccc|cccc|cccc|}
\hline
\multirow{7}{*}{DBSCAN} &  & \multicolumn{4}{c|}{PCA} & \multicolumn{4}{c|}{Kernel-PCA} & \multicolumn{4}{c|}{t-SNE} \\ \cline{2-14} 
 & \multirow{2}{*}{Sample Size} & \multicolumn{2}{c|}{raw features} & \multicolumn{2}{c|}{Q-transformed} & \multicolumn{2}{c|}{raw features} & \multicolumn{2}{c|}{Q-transformed} & \multicolumn{2}{c|}{raw features} & \multicolumn{2}{c|}{Q-tranformed} \\ \cline{3-14} 
 &  & \multicolumn{1}{c|}{K} & \multicolumn{1}{c|}{ARI} & \multicolumn{1}{c|}{K} & ARI & \multicolumn{1}{c|}{K} & \multicolumn{1}{c|}{ARI} & \multicolumn{1}{c|}{K} & ARI & \multicolumn{1}{c|}{K} & \multicolumn{1}{c|}{ARI} & \multicolumn{1}{c|}{K} & ARI \\ \cline{2-14} 
 & 50 & \multicolumn{1}{c|}{1} & \multicolumn{1}{c|}{0} & \multicolumn{1}{c|}{4} & 0.63 & \multicolumn{1}{c|}{1} & \multicolumn{1}{c|}{0} & \multicolumn{1}{c|}{4} & 0.64 & \multicolumn{1}{c|}{3} & \multicolumn{1}{c|}{0.52} & \multicolumn{1}{c|}{4} & 0.96 \\ \cline{2-14} 
 & 40 & \multicolumn{1}{c|}{1} & \multicolumn{1}{c|}{0} & \multicolumn{1}{c|}{4} & 0.60 & \multicolumn{1}{c|}{1} & \multicolumn{1}{c|}{0} & \multicolumn{1}{c|}{4} & 0.64 & \multicolumn{1}{c|}{3} & \multicolumn{1}{c|}{0.57} & \multicolumn{1}{c|}{4} & 0.98 \\ \cline{2-14} 
 & 30 & \multicolumn{1}{c|}{1} & \multicolumn{1}{c|}{0} & \multicolumn{1}{c|}{4} & 0.60 & \multicolumn{1}{c|}{1} & \multicolumn{1}{c|}{0} & \multicolumn{1}{c|}{4} & 0.63 & \multicolumn{1}{c|}{3} & \multicolumn{1}{c|}{0.60} & \multicolumn{1}{c|}{4} & 0.91 \\ \cline{2-14} 
 & 20 & \multicolumn{1}{c|}{1} & \multicolumn{1}{c|}{0} & \multicolumn{1}{c|}{4} & 0.71 & \multicolumn{1}{c|}{1} & \multicolumn{1}{c|}{0} & \multicolumn{1}{c|}{4} & 0.58 & \multicolumn{1}{c|}{3} & \multicolumn{1}{c|}{0.52} & \multicolumn{1}{c|}{4} & 0.87 \\ \hline
\multirow{4}{*}{GMM} & 50 & \multicolumn{1}{c|}{5} & \multicolumn{1}{c|}{0.53} & \multicolumn{1}{c|}{4} & 0.84 & \multicolumn{1}{c|}{5} & \multicolumn{1}{c|}{0.46} & \multicolumn{1}{c|}{4} & 0.79 & \multicolumn{1}{c|}{4} & \multicolumn{1}{c|}{0.72} & \multicolumn{1}{c|}{4} & 0.97 \\ \cline{2-14} 
 & 40 & \multicolumn{1}{c|}{5} & \multicolumn{1}{c|}{0.54} & \multicolumn{1}{c|}{4} & 0.83 & \multicolumn{1}{c|}{5} & \multicolumn{1}{c|}{0.54} & \multicolumn{1}{c|}{4} & 0.62 & \multicolumn{1}{c|}{3} & \multicolumn{1}{c|}{0.57} & \multicolumn{1}{c|}{4} & 0.98 \\ \cline{2-14} 
 & 30 & \multicolumn{1}{c|}{4} & \multicolumn{1}{c|}{0.53} & \multicolumn{1}{c|}{4} & 0.79 & \multicolumn{1}{c|}{4} & \multicolumn{1}{c|}{0.55} & \multicolumn{1}{c|}{4} & 0.79 & \multicolumn{1}{c|}{4} & \multicolumn{1}{c|}{0.74} & \multicolumn{1}{c|}{4} & 1 \\ \cline{2-14} 
 & 20 & \multicolumn{1}{c|}{5} & \multicolumn{1}{c|}{0.54} & \multicolumn{1}{c|}{4} & 0.93 & \multicolumn{1}{c|}{5} & \multicolumn{1}{c|}{0.57} & \multicolumn{1}{c|}{4} & 0.66 & \multicolumn{1}{c|}{5} & \multicolumn{1}{c|}{0.63} & \multicolumn{1}{c|}{4} & 0.96 \\ \hline
\multirow{4}{*}{X-means} & 50 & \multicolumn{1}{c|}{1} & \multicolumn{1}{c|}{0} & \multicolumn{1}{c|}{4} & \textbf{0.86} & \multicolumn{1}{c|}{1} & \multicolumn{1}{c|}{0} & \multicolumn{1}{c|}{4} & \textbf{0.87} & \multicolumn{1}{c|}{7} & \multicolumn{1}{c|}{0.63} & \multicolumn{1}{c|}{4} & 0.97 \\ \cline{2-14} 
 & 40 & \multicolumn{1}{c|}{1} & \multicolumn{1}{c|}{0} & \multicolumn{1}{c|}{4} & \textbf{0.87} & \multicolumn{1}{c|}{2} & \multicolumn{1}{c|}{0.26} & \multicolumn{1}{c|}{4} & \textbf{0.81} & \multicolumn{1}{c|}{7} & \multicolumn{1}{c|}{0.64} & \multicolumn{1}{c|}{4} & 0.98 \\ \cline{2-14} 
 & 30 & \multicolumn{1}{c|}{1} & \multicolumn{1}{c|}{0} & \multicolumn{1}{c|}{4} & \textbf{0.81} & \multicolumn{1}{c|}{6} & \multicolumn{1}{c|}{0.52} & \multicolumn{1}{c|}{4} & \textbf{0.79} & \multicolumn{1}{c|}{6} & \multicolumn{1}{c|}{0.68} & \multicolumn{1}{c|}{4} & 1 \\ \cline{2-14} 
 & 20 & \multicolumn{1}{c|}{1} & \multicolumn{1}{c|}{0} & \multicolumn{1}{c|}{4} & \textbf{0.87} & \multicolumn{1}{c|}{1} & \multicolumn{1}{c|}{0} & \multicolumn{1}{c|}{4} & \textbf{0.89} & \multicolumn{1}{c|}{3} & \multicolumn{1}{c|}{0.52} & \multicolumn{1}{c|}{4} & 0.93 \\ \hline
\end{tabular}
\end{table*}

\begin{table*}[!ht]
\caption{X-means based clustering (MOBIKEY dataset). K stands for an identified number of clusters. ARI stands for the adjusted rand index. X-means predicts the number of clusters successfully in quantile transformed space. Larger ARI values in the table below show that X-means outperforms both DBSCAN and GMM. }
\label{table:xMeans_Mobikey}
\centering
\begin{tabular}{|c|c|c|c|c|c|c|c|c|c|c|c|c|}
\hline
   & \multicolumn{4}{c|}{PCA} & \multicolumn{4}{c|}{Kernel-PCA} & \multicolumn{4}{c|}{t-SNE} \\ \hline
\multirow{2}{*}{Sample Size} &
  \multicolumn{2}{c|}{raw features} &
  \multicolumn{2}{c|}{Q-transformed} &
  \multicolumn{2}{c|}{raw features} &
  \multicolumn{2}{c|}{Q-transformed} &
  \multicolumn{2}{c|}{raw features} &
  \multicolumn{2}{c|}{Q-transformed} \\ \cline{2-13} 
   & K   & ARI      & K  & ARI   & K    & ARI     & K    & ARI     & K   & ARI    & K   & ARI   \\ \hline
50 & 2   & 0.34     & 4  & 0.59  & 5    & 0.30    & 4    & 0.62    & 4   & 0.61   & 4   & 0.69  \\ \hline
40 & 3   & 0.63     & 4  & 0.63  & 5    & 0.313   & 4    & 0.69    & 5   & 0.46   & 4   & 0.77  \\ \hline
30 & 10  & 0.26     & 4  & 0.68  & 9    & 0.25    & 4    & 0.79    & 5   & 0.49   & 4   & 0.83     \\ \hline
20 & 4   & 0.64     & 4  & 0.82  & 4    & 0.65    & 4    & 0.83    & 3   & 0.67   & 4   & 0.84  \\ \hline
\end{tabular}
\end{table*}

The quantitative scores based on ARI values are summarized in Table \ref{table:DBSCAN_GMM_Xmeans}. DBSCAN did not perform well with raw keystroke features. However, it can predict the number of clusters successfully with the quantile transformed data. ARI indices are marginally better for Kernel-PCA compared to PCA. However, ARI indices for t-SNE are better than both PCA and Kernel-PCA. 

In a next evaluation, GMM clustering is applied to the data. The BIC score is used to predict the number of clusters in the data (see Section \ref{para:xmeans} for the definition of BIC score) 
ARI is used to evaluate the results, which are tabulated in Table \ref{table:DBSCAN_GMM_Xmeans}. GMM predicted a higher number of clusters when operating on raw keystroke features, but estimated the correct number when applied to the quantile-transformed data and yielded higher ARI scores. Furthermore, the ARI scores followed similar trends to those seen with DBSCAN, with t-SNE outperforming Kernel PCA, and both outperforming PCA. 

Finally, X-means clustering is applied to the reduced  projected features. X-means can successfully estimate number of clusters in quantile transformed feature space with a higher ARI index than both GMM and DBSCAN. ARI index improves as we move from PCA to Kernel-PCA to t-SNE due to better separation of features (see Figure \ref{fig:quantile transformation} and Figure \ref{fig:quantile_transformation_mobiKey}).  

We observed a similar trend in the MOBIKEY dataset with DBSCAN, GMM, and X-means algorithms. The result of the X-means clustering algorithm is shown in Table \ref{table:xMeans_Mobikey}. As the MOBIKEY data is collected from a tablet, the touch data is susceptible to extra noise, leading to lower ARI scores than those seen with keyboard data. Here too, the quantile transform helps in improving the separation and clustering of incoming features. Also, as can be seen from the table, by using the most recent 20 incoming features, we can cluster users more accurately. As we include more samples from the past, the ARI score drops due to behavioral changes in the typing pattern of users over long periods of time \cite{ayotte2021study}. Overall, we found X-means to perform the best among the investigated clustering methods.

\begin{table*}[!ht]
\caption{Identification accuracy (CMU dataset): Identification accuracy is estimated for different sample sizes for the CMU keystroke dataset. We selected four users from the final session. For a ``sample size of $N$'', $0.8N$ samples are randomly selected for training, and $0.2N$ samples for testing as discussed in Section \ref{ssec:classification_acc}. First, identification accuracy for a sample size of 50 is estimated. Overall, accuracy with Ordinal UNLOC is slightly lower compared to PCA with the nearest neighbor approach. However, as we reduce the sample size, the accuracy of the proposed method improves. With a sample size of 20, the proposed method performs better for PCA and Kernel-PCA when compared to t-SNE. As we move from PCA space to Kernel-PCA space, the accuracy improves. However, tuning Kernel-PCA with the incoming feature vector is an open challenge. Features are clustered and separated well in the t-SNE domain. Hence, the accuracy with t-SNE is higher. However, since the embeddings in the t-SNE domain change in every run, the distance between points in the lower dimensional embedding is not always the same. Therefore, we observe a drop in accuracy when we use the proposed method Ordinal UNLOC with t-SNE.}
\label{table:CMU classification}
\centering
\resizebox{2\columnwidth}{!}{\begin{tabular}{|c|c|l|c|l|c|l|}
\hline
\multirow{2}{*}{Sample Size} &
  \begin{tabular}[c]{@{}c@{}}PCA, \\ nearest neighbor\end{tabular} &
  \begin{tabular}[c]{@{}l@{}}Ordinal UNLOC  \\ PCA, \\ nearest neighbor\end{tabular} &
  \begin{tabular}[c]{@{}c@{}}Kernel-PCA,\\ nearest neighbor\end{tabular} &
  \begin{tabular}[c]{@{}l@{}}Ordinal UNLOC \\  KPCA,\\ nearest neighbor\end{tabular} &
  \begin{tabular}[c]{@{}c@{}}t-SNE, \\ nearest neighbor\end{tabular} &
  \begin{tabular}[c]{@{}l@{}}Ordinal UNLOC \\  t-SNE,\\ nearest neighbor\end{tabular} \\ \cline{2-7} 
  & Q-transformed & Q-transformed & Q-transformed & Q-transformed & Q-transformed & Q-transformed \\ \hline
50 & 94.04         & 93.81          & 94            & 93.67         & 98.87         & 92.79         \\ \hline
40 & 94.01         & 93.38         & 93.41         & 89.12         & 98.80         & 93.87         \\ \hline
30 & 91.84         & 92.5          & 91.21         & 87.12         & \cellcolor{blue!25}99.82         & 91.82         \\ \hline
20 & 93.06         & \cellcolor{blue!25}95.71          & 93.22         & \cellcolor{blue!25}94.21         & 98.32         & 93.34         \\ \hline
10 & 88.68         & 93.27          & 89.25         & 93.17         & 89.78         & \cellcolor{blue!25}95.65         \\ \hline
\end{tabular}}
\end{table*}

\begin{table*}[ht]
\caption{Identification accuracy (MOBIKEY results): The identification accuracy of our algorithms is evaluated for the MOBIKEY dataset. The users's up-down digraph is fed as input. For a ``sample size of $N$'', $0.8N$ samples are randomly selected for training, and $0.2N$ samples for testing as discussed in Section \ref{ssec:classification_acc}. Overall, accuracy with Ordinal UNLOC is slightly lower compared to PCA with the nearest neighbor approach. However, as we reduce the sample size, the accuracy of the proposed method improves. With a sample size of 10, the proposed method performs better for PCA and Kernel-PCA when compared to t-SNE. The results get better in the Kernel-PCA space. }
\label{table:MobiKey classification}
\centering
\resizebox{2\columnwidth}{!}{\begin{tabular}{|c|c|l|c|l|c|l|}
\hline
\multirow{2}{*}{Sample Size} &
  \begin{tabular}[c]{@{}c@{}}PCA, \\ nearest neighbor\end{tabular} &
  \begin{tabular}[c]{@{}l@{}}Ordinal UNLOC  \\ PCA, \\ nearest neighbor\end{tabular} &
  \begin{tabular}[c]{@{}c@{}}Kernel-PCA,\\ nearest neighbor\end{tabular} &
  \begin{tabular}[c]{@{}l@{}}Ordinal UNLOC \\  KPCA,\\ nearest neighbor\end{tabular} &
  \begin{tabular}[c]{@{}c@{}}t-SNE, \\ nearest neighbor\end{tabular} &
  \begin{tabular}[c]{@{}l@{}}Ordinal UNLOC \\  t-SNE,\\ nearest neighbor\end{tabular} \\ \cline{2-7} 
   & Q-transformed & Q-transformed & Q-transformed & Q-transformed & Q-transformed & Q-transformed \\ \hline
50 & 86.88         & \cellcolor{blue!25}88.66         & \cellcolor{blue!25} 88.68         & 87.12         & 95.58         & 92.79         \\ \hline
40 & 83.25         & 84.28         & 85.12         & 84.21         & 94.62         & 92.87         \\ \hline
30 & 83.75         & 85.67         & 84.24         & 83.76         & 93.13         & 92.28         \\ \hline
20 & 76.99         & 78.75         & 78.36         & 78.12         & 84.06         & \cellcolor{blue!25}85.34         \\ \hline
10 & 75.68         & 82.27          & 79.25         & 83.17         & 79.78         & \cellcolor{blue!25}84.65         \\ \hline
\end{tabular}}
\end{table*}

\begin{table*}[ht]
\caption{The identification accuracy of our algorithms is evaluated by varying the number of multiple users accessing the system for the CMU dataset. Here, we carried out the experiment assuming no knowledge of number of users and their samples in the training dataset. We picked random sample size with an Monte-Carlo experiment on intra and inter session data.  We observe that the accuracy improves as we move from PCA to Kernel PCA to t-SNE. However, as the number of users accessing the system increases, the identification accuracy drops.}
\label{table:Multi user scenario}
\centering
\resizebox{2\columnwidth}{!}{\begin{tabular}{|l|l|ll|ll|ll|ll|ll|ll|}
\hline
\multirow{2}{*}{\begin{tabular}[c]{@{}l@{}}Number \\ of \\ Users\end{tabular}} & \multirow{2}{*}{Dimension} & \multicolumn{2}{l|}{\begin{tabular}[c]{@{}l@{}}PCA,\\ nearest neighbor\end{tabular}} & \multicolumn{2}{l|}{\begin{tabular}[c]{@{}l@{}}Ordinal UNLOC,\\ PCA, nearest\\ neighbor\end{tabular}} & \multicolumn{2}{l|}{\begin{tabular}[c]{@{}l@{}}KPCA,\\ nearest neighbor\end{tabular}} & \multicolumn{2}{l|}{\begin{tabular}[c]{@{}l@{}}Ordinal UNLOC,\\ KPCA, nearest \\ neighbor\end{tabular}} & \multicolumn{2}{l|}{\begin{tabular}[c]{@{}l@{}}t-SNE,\\ nearest neighbor\end{tabular}} & \multicolumn{2}{l|}{\begin{tabular}[c]{@{}l@{}}Ordinal UNLOC,\\ t-SNE, nearest \\ neighbor\end{tabular}} \\ \cline{3-14} 
 &  & \multicolumn{1}{l|}{\begin{tabular}[c]{@{}l@{}}Intra-\\ Session\end{tabular}} & \begin{tabular}[c]{@{}l@{}}Inter-\\ Session\end{tabular} & \multicolumn{1}{l|}{\begin{tabular}[c]{@{}l@{}}Intra-\\ Session\end{tabular}} & \begin{tabular}[c]{@{}l@{}}Inter-\\ Session\end{tabular} & \multicolumn{1}{l|}{\begin{tabular}[c]{@{}l@{}}Intra-\\ Session\end{tabular}} & \begin{tabular}[c]{@{}l@{}}Inter-\\ Session\end{tabular} & \multicolumn{1}{l|}{\begin{tabular}[c]{@{}l@{}}Intra-\\ Session\end{tabular}} & \begin{tabular}[c]{@{}l@{}}Inter-\\ Session\end{tabular} & \multicolumn{1}{l|}{\begin{tabular}[c]{@{}l@{}}Intra-\\ Session\end{tabular}} & \begin{tabular}[c]{@{}l@{}}Inter-\\ Session\end{tabular} & \multicolumn{1}{l|}{\begin{tabular}[c]{@{}l@{}}Intra-\\ Session\end{tabular}} & \begin{tabular}[c]{@{}l@{}}Inter-\\ Session\end{tabular} \\ \hline
3 & 2 & \multicolumn{1}{l|}{91.09} & 83.95 & \multicolumn{1}{l|}{96.67} & 85.11 & \multicolumn{1}{l|}{91.88} & 85.12 & \multicolumn{1}{l|}{97.02} & 86.67 & \multicolumn{1}{l|}{98.64} & 88.12 & \multicolumn{1}{l|}{92.21} & 82.72 \\ \hline
4 & 3 & \multicolumn{1}{l|}{91.15} & 83.45 & \multicolumn{1}{l|}{96.12} & 86.42 & \multicolumn{1}{l|}{92.01} & 84.61 & \multicolumn{1}{l|}{97.42} & 87.64 & \multicolumn{1}{l|}{98.22} & 88.21 & \multicolumn{1}{l|}{93.44} & 83.42 \\ \hline
5 & 4 & \multicolumn{1}{l|}{86.29} & 83.11 & \multicolumn{1}{l|}{96.32} & 85.81 & \multicolumn{1}{l|}{85.91} & 84.41 & \multicolumn{1}{l|}{96.67} & 86.21 & \multicolumn{1}{l|}{95.11} & 86.44 & \multicolumn{1}{l|}{90.12} & 80.21 \\ \hline
6 & 4 & \multicolumn{1}{l|}{72.21} & 71.01 & \multicolumn{1}{l|}{76.80} & 76.58 & \multicolumn{1}{l|}{73.14} & 72.12 & \multicolumn{1}{l|}{78.12} & 76.44 & \multicolumn{1}{l|}{94.92} & 78.22 & \multicolumn{1}{l|}{80.12} & 78.48 \\ \hline
\end{tabular}}
\end{table*}

\subsection{Classification accuracy}
\label{ssec:classification_acc}
First, we select four random  users from CMU keystroke dataset. In the CMU keystroke dataset, each user has a total of 400 samples taken from 8 different sessions, with each user typing a password 50 times during a session. For each user, the samples are selected randomly from all the sessions, outliers are removed from the dataset, and an 80/20 percent split is performed on the data to separate the points into training and testing sets. Specifically, when the result is reported for a ``sample size of $N$'', $0.8N$ samples are randomly selected for training, and $0.2N$ samples for testing. Our proposed algorithm (see Section \ref{sec:new_algo}) is then applied to this data. Anchor locations are derived as described in the training phase, and the data in the testing set are treated as keystrokes from ``unknown'' users. Once the locations of the ``unknown'' users are estimated, they are classified based on the nearest neighbor rule. To benchmark the performance of the proposed algorithm, the test features are also classified by just the nearest neighbor rule in the reduced feature spaces of PCA, Kernel-PCA, and t-SNE without using Ordinal UNLOC.

Identification accuracy is estimated for different sample sizes for the CMU keystroke dataset. As it is difficult to identify number of user clusters from the raw input data (see Table \ref{table:DBSCAN_GMM_Xmeans} and \ref{table:xMeans_Mobikey}), the pre-processing steps discussed in section \ref{ssec:effect_quantile} are crucial to the performance of the classification algorithm. With unprocessed data, the numbers of clusters are frequently wrongly estimated, leading to poor performance (<25\% accuracy), and when the number of clusters is correctly estimated, classification on unprocessed data leads to better performance ($\sim$ 89\% accuracy), but still does not match the performance when the algorithm is applied to processed data (see Table \ref{table:CMU classification}).  Overall, accuracy with Ordinal UNLOC is slightly lower compared to PCA with the nearest neighbor approach. However, as we reduce the sample size, the accuracy of our proposed algorithm improves. As can be seen in Table \ref{table:CMU classification},  for a sample size of 10, it can be seen that Ordinal UNLOC when applied after  PCA or Kernel-PCA, outperform other methods. This is noteworthy, since accurate identification is possible with as few as 10 keystroke samples using our approach - making it a candidate for use alongside password entry systems. 

As we move from PCA space to Kernel-PCA space, the accuracy improves. However, tuning Kernel-PCA with the incoming feature vector is a challenge. Features are clustered and separated well in the t-SNE domain. Hence, the accuracy with t-SNE is higher. However, since embeddings in the t-SNE domain change in every run, the distances between features in the lower-dimensional embedding do not stay the same as the data changes. This leads to a drop in overall accuracy.

To further demonstrate the effectiveness of our approach, we carried out intra-session and inter-session analysis on the CMU-Keystroke dataset (Table \ref{table:Multi user scenario}). Further, we assumed no prior knowledge of the number of users and their sample sizes in the training and test sets. The identification accuracy of our algorithms is evaluated by varying the number of multiple users accessing the system for the CMU dataset. We performed Monte-Carlo experiments with random sample sizes and intra and inter session data. To improve separability among user feature clouds, the number of dimensions in reduced feature space of PCA/ Kernel-PCA is varied. The lowest number of dimensions needed to show significant performance gains and the corresponding identification accuracy values are shown in Table \ref{table:Multi user scenario}. We observe that the accuracy improves as we move from PCA to Kernel PCA to t-SNE. However, as the number of users accessing the system increases, the identification accuracy drops. As the samples are drawn randomly in both intra and inter sessions, we observe the following: 
\begin{itemize}
    \item A drop in identification performance in classical approaches (PCA/KPCA/t-SNE with nearest neighbor) without ordinal ULOC. The intra session performance further dropped by 5\% or more.
    \item The intra session performance is agnostic to random sample selection with ordinal UNLOC. However, with inter session data, the performance drops due to the effect of time in behavioral keystroke biometrics \cite{ayotte2020fast}.
\end{itemize}

The accuracy of our algorithms is evaluated for the MOBIKEY dataset \cite{antal2016mobikey} as well. The touch up - touch down features are used as input. In the MOBIKEY dataset, there are 20 samples per session and a total of 3 sessions. We randomized the samples from 3 sessions to obtain the results. Hence, they represent a generalization of the inter session performance, and the results are shown in Table \ref{table:MobiKey classification}. The results are compared with classical approaches. We observed, again, that the proposed method outperforms the classical methods. Other observations are consistent with what was seen with the CMU dataset.

Furthermore, as the number of users accessing the system increases, the identification accuracy drops gradually. In all cases, our proposed algorithm, combined with PCA/Kernel-PCA space performs marginally better compared to just PCA/Kernel-PCA with the nearest neighbor approach.

\section{Conclusions}
\label{sec:conclusions}

In this paper, we presented an algorithm to solve a multi-user identification problem in keystroke dynamics. Quantile transformed keystroke samples are projected to a lower dimensional space using PCA/ Kernel-PCA, or t-SNE. Ordinal-UNLOC is applied to an incoming quantile transformed sample in the reduced feature space to find its optimal ``location'' in the reduced feature space, and based on this location, the test sample is classified based on the nearest neighbor rule. 

To test the effectiveness of our proposed algorithm, classification accuracy is estimated for different sample sizes and compared with traditional PCA/Kernel-PCA/t-SNE with the nearest neighbor classifier, using data from the CMU dataset and the MOBIKEY dataset. Among all the algorithms, our proposed algorithm is  effective in identifying multiple users.

Keystroke biometrics is a behavioral biometric modality, where keystroke patterns continuously evolve with time. Hence, including more samples from the past deteriorates the performance metrics \cite{ayotte2020fast}. Future work includes the investigation of typing behavior with time, and how that can affect the performance of this and similar identification systems. Future work also includes implementing a deep learning framework for multi-user classification, the investigation of typing behavior with time, and how that can affect the performance of this and similar classification systems. Further exploration is warranted to modify this algorithm for use in datasets such as Clarkson Account Recovery dataset \cite{wahab2022securing}. Other distance measures such as the Mahalanobis distance can be investigated for their effectiveness when used as part of the proposed algorithm.

\section*{Acknowledgments}

This work was supported in part by NSF CPS 1646542.

\ifCLASSOPTIONcaptionsoff
  \newpage
\fi



%
\bibliographystyle{IEEEtran}
\bibliography{main.bib}
%

\begin{IEEEbiography}[{\includegraphics[width=1in,height=1.25in,clip,keepaspectratio]{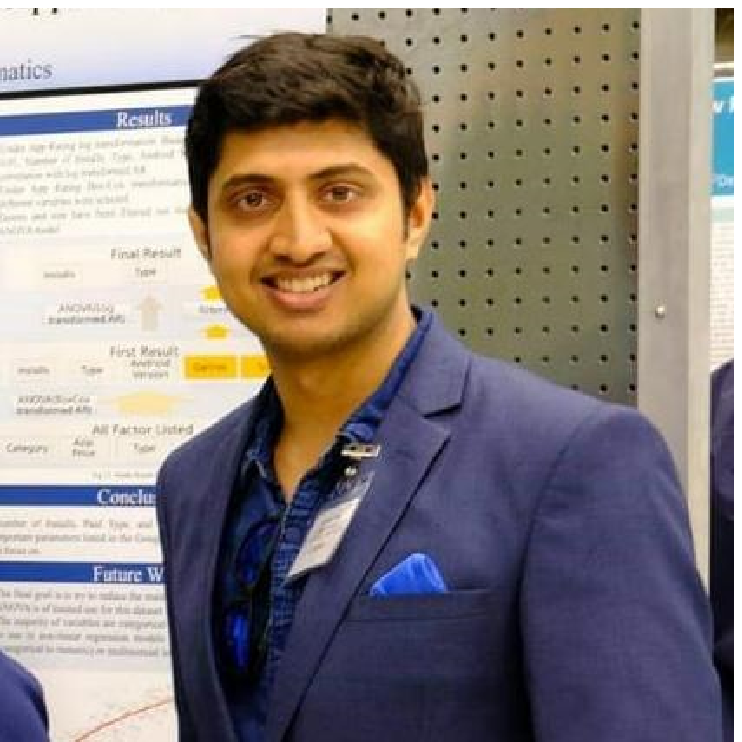}}]{Chinmay Sahu}
received the B.Tech. degree from Biju Patnaik University of Technology, Odisha, India, and the M.S. from the National Institute of Technology, Tiruchirappalli, Tamilnadu, India. Before joining Clarkson University, he worked as a software designer with Alstom Transport India Ltd. Currently, his research interest includes designing and developing algorithms to solve problems in fields related to the Internet of Things (IoT), biomedical engineering, geo-hazards, behavioral biometrics.
\end{IEEEbiography}

\begin{IEEEbiography}[{\includegraphics[width=1in,height=1.25in,clip,keepaspectratio]{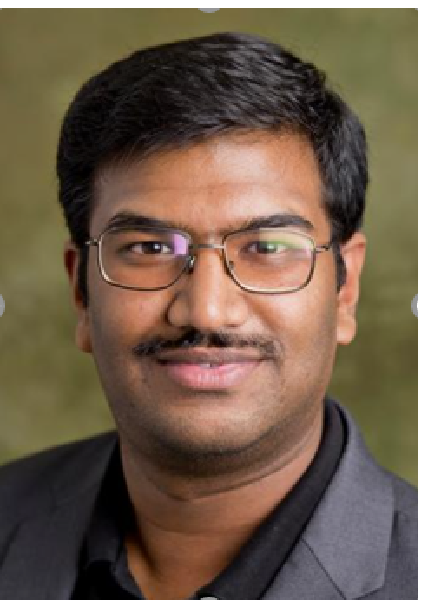}}]{Mahesh Banavar} received the B.E. degree in telecommunications engineering from Visvesvaraya Technological University, in 2005, and the M.S. and Ph.D. degrees in electrical engineering from Arizona State University, in 2007 and 2010, respectively. He is currently an Associate Professor with the Department of ECE, Clarkson University, Potsdam, NY, USA. His interests include node localization, detection and estimation algorithms, and user-behavior-based cybersecurity applications.
\end{IEEEbiography}
\vfill
\begin{IEEEbiography}[{\includegraphics[width=1in,height=1.25in,clip,keepaspectratio]{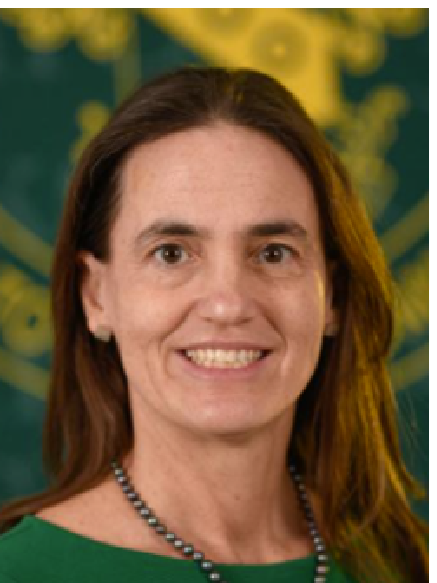}}]{Stephanie Schuckers} is the Paynter-Krigman
Endowed Professor in Engineering Science in
the Department of Electrical and Computer Engineering at Clarkson University and serves as
the Director of the Center for Identification Technology Research (CITeR), a National Science
Foundation Industry/University Cooperative Research Center. She received her doctoral degree in Electrical Engineering from The University of Michigan. Professor Schuckers research focuses on processing and interpreting signals which arise from the human body. Her work is funded from various sources, including National Science Foundation, Department of Homeland Security, and private industry, among others. She has started her own business, testified for US Congress, and has over 40 journal publications as well as over 60 other academic publications.
\end{IEEEbiography}
\flushbottom



\end{document}